# Title: Exciton Fission Enhanced Silicon Solar Cell

**Authors:** Narumi Nagaya[1,2]†, Kangmin Lee[1]†, Collin F. Perkinson[1,3]†, Aaron Li[1,3], Youri Lee[4], Xinjue Zhong[5], Sujin Lee[5], Leah P. Weisburn[3], Tomi K. Baikie[1], Moungi G. Bawendi[3], Troy Van Voorhis[3], William A. Tisdale[2], Antoine Kahn[5], Kwanyong Seo[4], Marc A. Baldo*[1]

**Affiliations:**

[1]Research Laboratory of Electronics, Massachusetts Institute of Technology; Cambridge, Massachusetts 02139, USA

[2]Department of Chemical Engineering, Massachusetts Institute of Technology; Cambridge, Massachusetts 02139, USA

[3]Department of Chemistry, Massachusetts Institute of Technology; Cambridge, Massachusetts 02139, USA

[4]School of Energy and Chemical Engineering, Ulsan National Institute of Science and Technology (UNIST); Ulsan 44919, Korea

[5]Department of Electrical and Computer Engineering, Princeton University; Princeton, New Jersey 08544, USA

*Corresponding author. Email: baldo@mit.edu

†These authors contributed equally to this work

**Abstract:**

While silicon solar cells dominate global photovoltaic energy production, their continued improvement is hindered by the single junction limit. One potential solution is to use molecular singlet exciton fission to generate two electrons from each absorbed high-energy photon. We demonstrate that the long-standing challenge of coupling molecular excited states to silicon solar cells can be overcome using sequential charge transfer. Combining zinc phthalocyanine, aluminum oxide, and a shallow junction crystalline silicon microwire solar cell, the peak charge generation efficiency per photon absorbed in tetracene is (138 ± 6)%, comfortably surpassing the quantum efficiency limit for conventional silicon solar cells and establishing a new, scalable approach to low cost, high efficiency photovoltaics.

**Main Text:**

Single junction crystalline silicon solar cells are rapidly approaching their theoretical efficiency limit of 29% (*1*). As in Fig. 1A, however, there are substantial opportunities to improve performance if thermalization losses can be reduced in the visible and UV spectrum. For example, using singlet exciton fission in tetracene (Tc) to double the available carriers obtained from blue and green sunlight could increase the power conversion efficiency of a single-junction crystalline silicon solar cell to 35% (*2*, *3*).

Singlet exciton fission in Tc generates two spin-1 triplet excitons with energies of 1.25 eV from one spin-0 singlet exciton with an energy of 2.4 eV (*4*); see Fig. 1B. Spin conservation opposes the thermalization of a singlet into one triplet exciton, protecting the process against losses and contributing to a near ideal yield in many fission materials, including Tc (*4*). The triplet energy of Tc is just above that of the 1.1 eV band gap of crystalline silicon (*c*-Si), which is ideal for coupling to *c*-Si. However, transferring the energy from the two excitons formed from one high energy photon absorbed in Tc to *c*-Si and observing device enhancements has proven challenging (*5*). Depositing Tc directly onto hydrogen-terminated *c*-Si surfaces decreases the external quantum efficiency (EQE) of both silicon-Tc heterojunction solar cells and interdigitated back-contacted (IBC) solar cells (*6*, *7*). Inserting LiF spacers (*8*, *9*) and pyrene passivation layers (*10*) between the Tc and silicon also does not yield efficient triplet energy transfer. Introducing a thin interlayer of $HfO_xN_y$ between Tc and *c*-Si, however, shows sensitization of *c*-Si by triplet excitons, albeit without demonstrating an increase in the photocurrent of an IBC cell (*11*). The $HfO_xN_y$ interlayer is hypothesized to play a bifunctional role, acting to passivate some silicon surface defects whilst simultaneously enabling triplet exciton transfer (*11*).

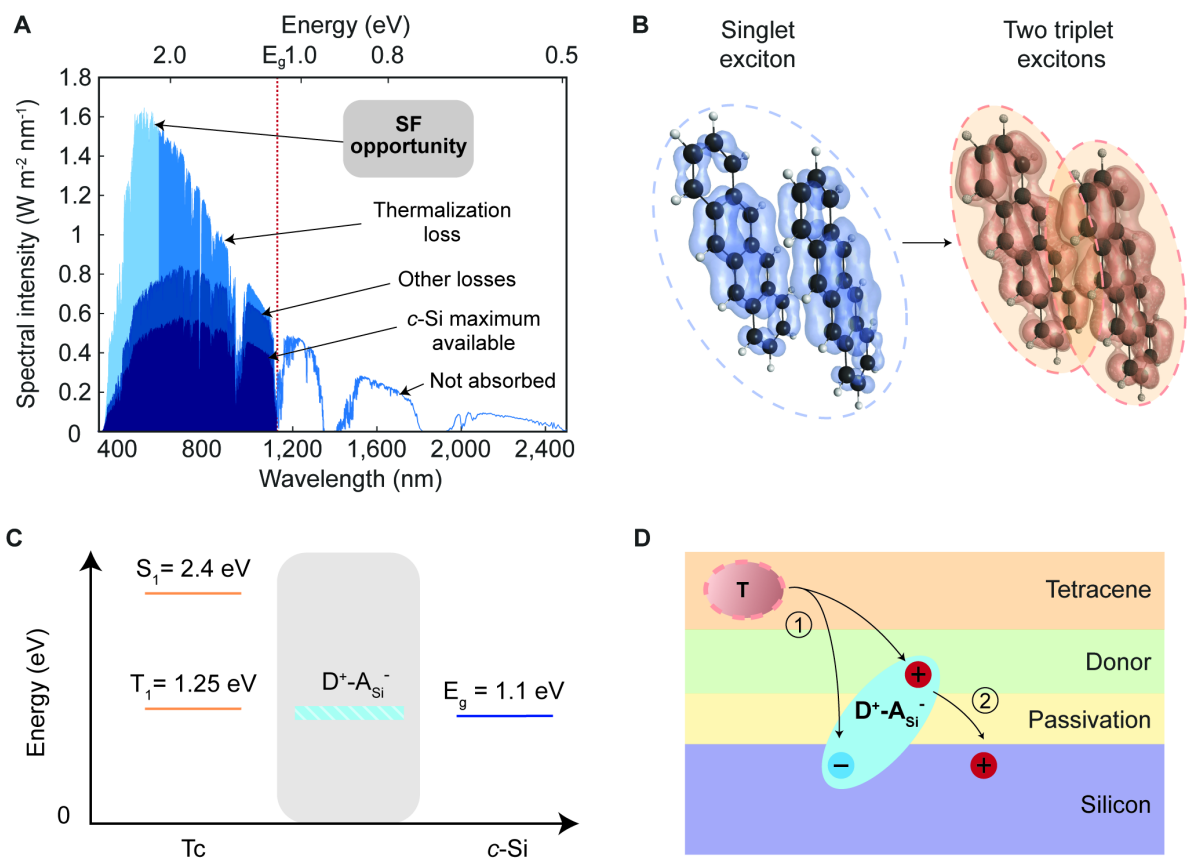

**Figure 1. Singlet exciton fission and a two-part interface design for coupling to silicon.** (**A**) The AM1.5G solar spectrum compared to the optical absorption range of crystalline silicon (*c*-Si). The red dotted line indicates the bandgap of *c*-Si. The portion of the spectrum to the right of the bandgap is not absorbed. To the left of the bandgap are spectral regions of the solar spectrum that are representative of energy pathways within a single junction silicon cell at the efficiency limit. These regions correspond to photons that are available to *c*-Si, other losses (*1*), the thermalization loss, and the opportunity for singlet fission-enhanced *c*-Si devices, respectively shaded from darkest to lightest blue. (**B**) A molecular picture of the singlet exciton fission process, showing the electron density of a delocalized singlet exciton over two Tc molecules in blue, and two product triplet excitons in orange. (**C**) Generalized energetic requirements for the donor-acceptor state ($D^+$-$A_{Si}^-$) for charge injection into silicon. The $D^+$-$A_{Si}^-$ energy should ideally lie between the triplet exciton energy of tetracene (Tc) and the bandgap of silicon. (**D**) The two-part interface design presented in this work. The interface consists of an electron-donating layer (green) and a passivation layer (yellow). The latter passivates silicon surface defects. We propose a sequential charge transfer mechanism: (1) the donor supports an initial electron transfer to silicon, forming $D^+$-$A_{Si}^-$ (light blue oval), followed by (2) a hole transfer from the donor to silicon, ultimately resulting in triplet energy transfer to silicon.

Here, we propose a two-component interface for efficient triplet exciton sensitization of *c*-Si. Our approach is guided by the apparent ineffectiveness of previous efforts to directly transfer triplets from Tc to c-Si via a Dexter mechanism (*8*, *9*). We instead design our cells based on the observed coupling between Tc and c-Si across thin $HfO_xN_y$ interfaces (*11*). We correlate evidence of triplet transfer with the presence of midgap defect states in $HfO_xN_y$, suggesting the possibility of sequential charge transfer via an intermediate charge-separated state supported by HfO$_x$N$_y$ (*12*). Consequently, we propose to mediate sensitization with an initial charge transfer to silicon (*11*, *13*), followed by a sequential transfer of the remaining charge carrier. The energy of the intermediate charge-separated state must lie between the triplet energy of Tc (1.25 eV) and the bandgap of silicon (1.1 eV). Tc itself is an electron donor (*14*), but at least in previously studied interfaces, its highest occupied molecular orbital (HOMO) is too deep to support a charge separated state within the necessary energy range. Therefore, we propose to insert an additional electron donor at the interface with silicon that supports a state D$^+$-A$_{Si}^-$ with the appropriate energy, where D$^+$ represents the charged electron donor and A$_{Si}^-$ describes the electron accepting role of silicon as in Fig. 1C.

The second component of our interface is a thin passivation layer necessary to prevent the transferred charge carriers from immediately recombining at the silicon surface, while still enabling carrier tunneling; see Fig. 1D. Aluminum oxide (AlO$_x$) is commonly used to passivate silicon solar cells (*15*). In this work, we use approximately 1-nm-thick layers of AlO$_x$ to both passivate the silicon surface and maintain charge tunneling across the interface.

**Zinc phthalocyanine as a donor material**

We study zinc phthalocyanine (ZnPc) as a candidate donor material at the Tc-Si interface. ZnPc is a common donor with a reported HOMO above that of Tc (*16*). Our time-dependent functional theory (TDDFT) calculations (see Supplementary Text) and literature (*17*) show that the excited Tc triplet state is 0.1-0.2 eV higher in energy than the ZnPc triplet, potentially enabling triplet harvesting at the interface (*18*) prior to the formation of a charge separated state. Further calculations show that the Tc HOMO is 0.2 eV deeper than the ZnPc HOMO.

We perform ultraviolet photoelectron spectroscopy to determine the energetic alignment of ZnPc at interfaces between Tc and *c*-Si. In Fig. 2A, we summarize the alignments for a highly n-doped Si surface (n$^+$-Si). The difference between the *c*-Si conduction band minimum and the HOMO level of ZnPc is approximately 1.20 eV, within the allowed energetic range for the D$^+$-A$_{Si}^-$ state, as depicted in Fig. 1C. We neglect binding energy in the determination of the energy of the charge-separated state D$_{ZnPc}^+$-A$_{Si}^-$ due to expected charge screening and delocalization in silicon (*19*, *20*). Furthermore, there is minimal barrier for subsequent hole transfer to the bulk valence band of silicon, within measurement error. We observe a negligible difference in the Tc-Si energy alignment with and without the ZnPc layer. This confirms that the presence of the ZnPc layer lowers the energy of a donor-acceptor state with silicon, thereby enabling sequential charge transfer to silicon following triplet exciton formation (see Fig. S1 and Supplementary Text). In contrast, the energetic alignment with a highly p-doped silicon surface (p$^+$-Si) shown in Fig. 2B exhibits a significant electron transfer barrier, and the charge separated state is the lowest energy state in the system, hindering overall energy transfer from Tc to silicon. Thus, we suspect that the energy levels of Tc/ZnPc enable sequential charge transfer to n$^+$-doped *c*-Si. The energetic alignment, however, is unfavorable for p$^+$-doped *c*-Si.

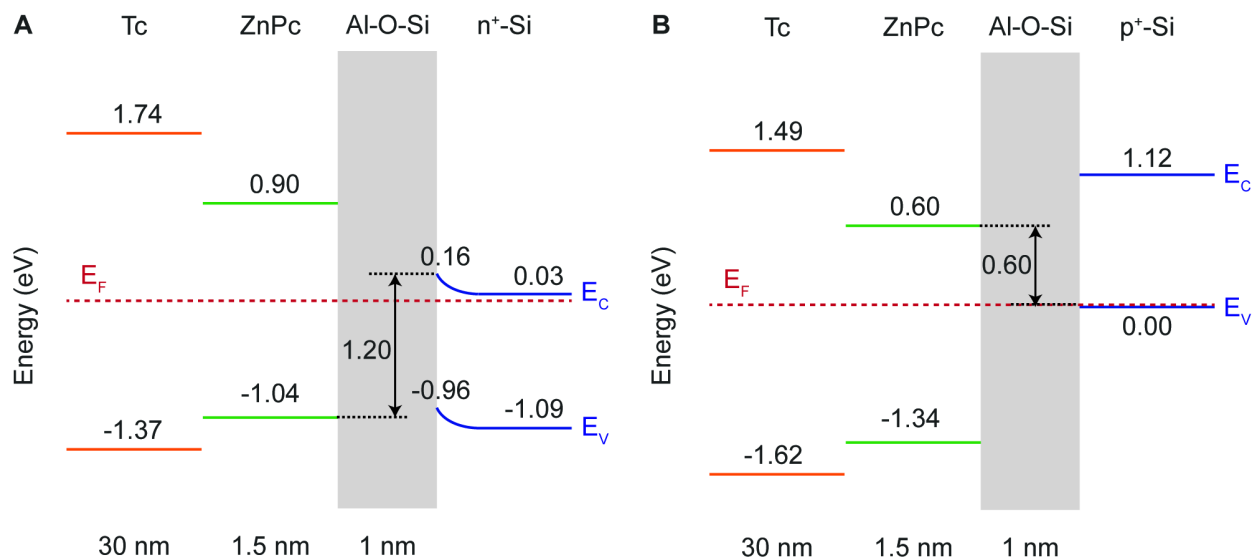

**Figure 2. Energetic alignments of Tc/ZnPc interfaces on n$^+$-doped and p$^+$-doped $c$-Si.**
Summary of band alignments for tetracene (Tc), zinc phthalocyanine (ZnPc) and aluminum oxide (AlO$_x$) deposited on (**A**) a highly n-doped silicon surface (n$^+$-Si), and (**B**) a highly p-doped silicon surface (p$^+$-Si). The red-dashed line indicates the position of the Fermi level (E$_F$) of the system. The valence band maximum (E$_V$) and highest occupied molecular orbital (HOMO) positions with respect to E$_F$ were measured using ultraviolet photoelectron spectroscopy (UPS). The conduction band minimum (E$_C$) and lowest unoccupied molecular orbital (LUMO) positions were calculated using the electronic band gaps from previous UPS and inverse photoemission spectroscopy measurements (*11*, *16*). The experimental energy resolution for measurements is typically 0.10 eV. The bulk E$_C$ and E$_V$ positions of the doped silicon were calculated from the doping concentration. The black-dashed lines show the energy of the lowest-lying state at the interface between the ZnPc layer and $c$-Si: in (A), the electron is located on E$_C$ and the hole is located on the HOMO of ZnPc; in (B), the electron is located on the LUMO of ZnPc and the hole is located on E$_V$. Nominal deposited thicknesses for the Tc, ZnPc and AlO$_x$ layers are listed.

**Silicon solar cell design for singlet fission sensitization**

Charge injection at the surface of silicon demands a solar cell design with efficient surface carrier collection. Here, we employ shallow p-n junctions, both in conventional planar geometries and microwire (MW) solar cells that employ shallow radial junctions. Shallow junctions efficiently extract charge carriers transferred to the surface of $c$-Si by reducing the propagation distance of minority carriers (*21*). The microwire structure also exhibits enhanced light absorption (*21*, *22*), particularly in the short wavelength region, reducing reflection losses.

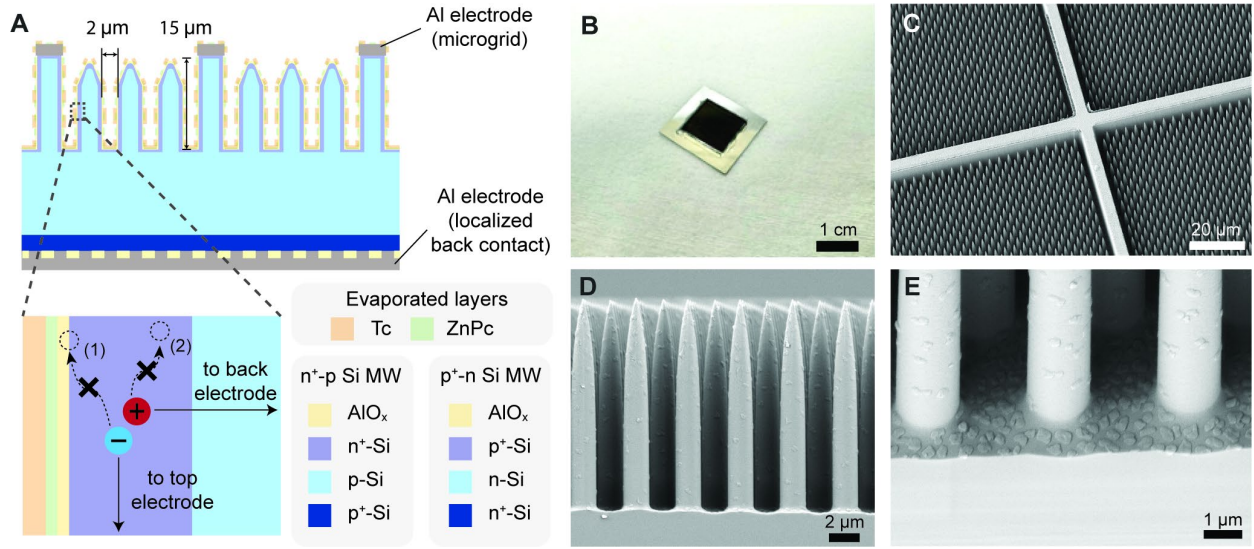

**Figure 3. Microwire c-Si solar cells.** (**A**) Schematic depicting layers of the microwire (MW) solar cells used in this work. MW inset shows that the solar cells are designed to extract carriers from the surface region of silicon: (1) Surface recombination losses are reduced by depositing an AlO$_x$ passivation layer; (2) bulk recombination losses are reduced by employing shallow radial p-n junctions through the microwire architecture, as well as a microgrid array for the top aluminum electrode. (**B**) Photograph of a fabricated n$^+$-p Si MW cell with deposited Tc and ZnPc layers, with encapsulating glass on the active area. (**C**) Scanning Electron Microscopy (SEM) image of the Tc/ZnPc/n$^+$-p Si MW cell showing a section of the microgrid array of the top electrode. (**D**) Transverse SEM image of the Tc/ZnPc/n$^+$-p Si MW cell, showing the deposition on the microwires. (**E**) SEM image of the Tc/ZnPc/n$^+$-p Si MW cell showing the deposition of the Tc/ZnPc layers focusing on the base of the microwire cell.

Figure 3A shows a schematic of the fabricated devices. Following a previously reported protocol (*23*, *24*), we fabricate tapered *c*-Si microwire arrays with a spacing of 2 μm and a length of 15 μm. A radial junction is formed on the p-type tapered c-Si microwire arrays with an n-type emitter junction depth of 300 nm (see Fig. S2) through a spin-on-dopant-based thermal doping process. To minimize recombination on the rear side, a back surface field (BSF) layer with a junction depth of 1 μm and a localized back contact is added. Then, a 1-nm-thick AlO$_x$ layer is formed using Atomic Layer Deposition (ALD) on the front side of the solar cell. A microgrid electrode is applied as the front electrode to efficiently collect carriers (*25*). This microgrid electrode covers only 2% of the active area (1 cm²), thereby reducing shading losses from the top electrodes, while also increasing the deposition area of Tc/ZnPc on the AlO$_x$/*c*-Si surface. Nominally 1.5 nm of ZnPc followed by 30 nm of Tc is then thermally evaporated onto the front surface of the microwire cells. For more details on the full device fabrication procedure, see the Materials and Methods section.

To investigate the role of ZnPc, we also prepare a control device through an identical method on a p$^+$-n device where we only swap the doping of the base microwire device structure (Fig. 3A). Figures 3D-E show scanning electron microscopy (SEM) images of the Tc/ZnPc/n$^+$-p Si MW. Island-type growth is observed on the sides of the MWs, with similar but denser growth on the bottom region of the microwire arrays. Since thermal evaporation is a directional deposition technique, increased deposition is expected to occur on the horizontal surface (bottom region) than on the sidewalls of the MWs perpendicular to the deposition sources (*26*).

**Device characterization**

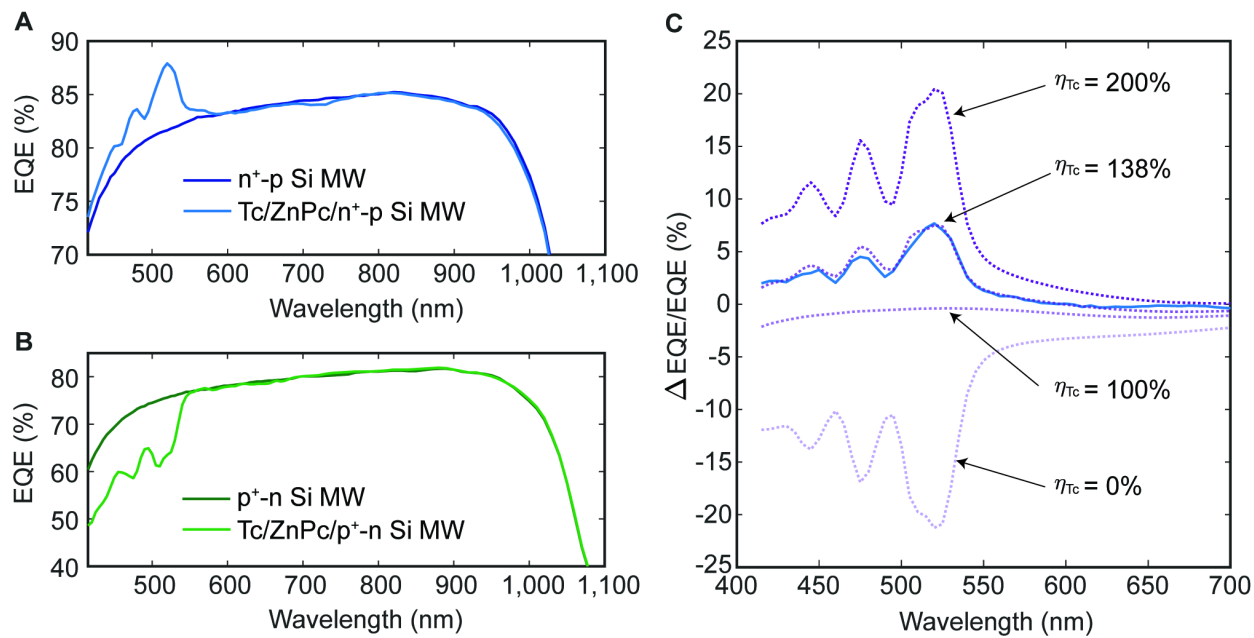

**Figure 4. Device External Quantum Efficiencies.** (**A**) Measured external quantum efficiency (EQE) spectra of $n^+$-p Si MW cells before and after Tc and ZnPc deposition. (**B**) Measured EQE spectra of $p^+$-n Si MW cells before and after Tc and ZnPc deposition. (**C**) Simulation fits of the percentage difference enhancement after organic deposition (ΔEQE/EQE) of the $n^+$-p Si MW cells presented in (A). The dotted lines represent simulated differential EQE at different Tc sensitization efficiencies ($\eta_{Tc}$). The solid line is data for the Tc/ZnPc/$n^+$-p Si MW device shown in (A).

The measured EQE spectra of the $n^+$-p MW device before and after Tc and ZnPc deposition are presented in Fig. 4A and Fig. S3B. After deposition of Tc/ZnPc, we see a positive contribution corresponding to the absorption spectrum of Tc, with a maximum EQE increase from 81.6% to 87.9% at 520 nm. The measured J-V curves of the devices also show that depositing ZnPc and Tc on the $n^+$-p MW devices results in an enhancement in the short-circuit current density, with negligible decrease in the open-circuit voltage and fill factors, resulting in an overall enhancement in power conversion efficiency, see Fig. S3 and Table S1. It is notable that the introduction of molecular materials does not degrade the electrical characteristics of the silicon solar cell. Our results suggest that the role of the molecular materials in this cell architecture is primarily excitonic in nature, and other than increased charge injection, decoupled from the operation of the junction itself.

As expected from the interfacial measurements summarized in Fig. 2, $p^+$-n MW devices exhibit dramatically different behavior when coupled to Tc. The measured EQE spectra before and after Tc and ZnPc deposition are shown in Fig. 4B and Fig. S4B. Unlike the case with the $n^+$-p MW device, we instead observe shadowing in the EQE spectra from absorption of Tc, dropping from 75.6% to 63.5% at 520 nm. This also confirms that the device performance enhancements observed in the $n^+$-p devices are not solely from enhanced antireflective effects of adding the additional organic layers.

We also fabricate n$^+$-p and p$^+$-n type planar devices with random surface pyramidal texturing, which is extensively utilized in the *c*-Si solar cell industry for its antireflective properties. The effects on the EQE spectra are consistent with the observations in the MW solar cells presented above, although the magnitude of the enhancement is slightly decreased (see Fig. S5B).

The role of charge tunneling in the formation of the D$^+$-A$_{Si}^-$ state is verified by replacing the 1-nm-thick AlO$_x$ passivation layer on an n$^+$-p MW solar cell with a 10 nm-thick Al$_2$O$_3$ passivation layer. On the thicker sample, we observe Tc shadowing in the EQE spectra, demonstrating that coupling to *c*-Si is disrupted (see Fig. S6A). We verify that ZnPc is essential by fabricating control devices without the ZnPc donor layer on both n$^+$-p and p$^+$-n MW devices, where we again observe Tc shadowing in the EQE spectra (see Fig. S6B-C). We note that while both n$^+$-p and p$^+$-n solar cells fabricated without the ZnPc layer exhibit shadowing from Tc, the shadowing is less in the n$^+$-p MW device. This suggests that there may be a small number of triplet states from Tc that are able to undergo Dexter transfer or sequential charge transfer to *c*-Si, potentially via charge-separated states incorporating the tail of the Tc density of states.

**Discussion**

The overall Tc sensitization efficiency, $\eta_{Tc}$, encompasses singlet fission efficiency in Tc, transport of triplets to the interface with silicon, triplet transfer efficiency at the interface, and extraction efficiency of charge carriers transferred to silicon. We emphasize that singlet exciton transfer alone from Tc to *c*-Si in the n$^+$-p MW device yields a maximum sensitization efficiency $\eta_{Tc}$ = 100%. Indeed, it is not possible to observe Tc peaks in the EQE of the silicon solar cell unless there is efficient ($\eta_{Tc}$ > 100%) coupling to triplets generated by singlet exciton fission. To evaluate $\eta_{Tc}$ we perform a transfer matrix model of the layers in our solar cell (*27*). Calculation details are presented in Supplementary Text. Figure 4C shows the fitting results to the EQE percentage differential (ΔEQE/EQE$_0$) where ΔEQE = EQE$_{Tc}$ – EQE$_0$ and EQE$_{Tc}$ and EQE$_0$ are the efficiencies of the Tc and control devices, respectively. The fit yields a sensitization efficiency of $\eta_{Tc}$ = (138 ± 6)%, where 200% is the maximum theoretical enhancement possible for a photocurrent doubling process.

Based on our results, we expect that it is possible to significantly improve the performance of singlet exciton fission enhanced solar cells and photodetectors toward a peak quantum yield of two electrons per photon and an overall power efficiency that exceeds the conventional single junction limit. First, technologies from advanced-node field-effect transistors can be adopted to improve the performance of the thin passivation layer. Second, conformal coating and optimized Tc growth could substantially improve the coverage of sensitizing materials on the silicon surface. Finally, although it has long been the archetype fission material for coupling to silicon, Tc has poor photostability (*28*) and will need to be replaced with a photostable analog.

In conclusion, we demonstrate efficient coupling between a silicon solar cell and singlet exciton fission in Tc, finally realizing the solar cell concept first proposed by D. L. Dexter in 1979[1]. Control over the crucial interface between Tc and silicon is established by assuming sequential charge transfer mediated by a thin layer of ZnPc. The resulting observation of more than one electron per photon in a silicon solar cell provides the foundation for a new solar photovoltaic technology capable of accelerating the global adoption of renewable energy.

**References and Notes**

1. W. Shockley, H. J. Queisser, Detailed Balance Limit of Efficiency of p-n Junction Solar Cells. *Journal of Applied Physics* **32**, 510–519 (1961).

2. D. L. Dexter, Two ideas on energy transfer phenomena: Ion-pair effects involving the OH stretching mode, and sensitization of photovoltaic cells. *Journal of Luminescence* **18–19**, 779–784 (1979).

3. A. Rao, R. H. Friend, Harnessing singlet exciton fission to break the Shockley–Queisser limit. *Nature Reviews Materials* **2**, 1–12 (2017).

4. M. B. Smith, J. Michl, Singlet Fission. *Chem. Rev.* **110**, 6891–6936 (2010).

5. A. J. Baldacchino, M. I. Collins, M. P. Nielsen, T. W. Schmidt, D. R. McCamey, M. J. Y. Tayebjee, Singlet fission photovoltaics: Progress and promising pathways. *Chemical Physics Reviews* **3**, 021304 (2022).

6. R. W. MacQueen, M. Liebhaber, J. Niederhausen, M. Mews, C. Gersmann, S. Jäckle, K. Jäger, M. J. Y. Tayebjee, T. W. Schmidt, B. Rech, K. Lips, Crystalline silicon solar cells with tetracene interlayers: the path to silicon-singlet fission heterojunction devices. *Mater. Horiz.* **5**, 1065–1075 (2018).

7. B. Daiber, S. Maiti, S. M. Ferro, J. Bodin, A. F. J. van den Boom, S. L. Luxembourg, S. Kinge, S. P. Pujari, H. Zuilhof, L. D. A. Siebbeles, B. Ehrler, Change in Tetracene Polymorphism Facilitates Triplet Transfer in Singlet Fission-Sensitized Silicon Solar Cells. *J. Phys. Chem. Lett.*, 8703–8709 (2020).

8. T. Hayashi, T. G. Castner, R. W. Boyd, Quenching of molecular fluorescence near the surface of a semiconductor. *Chemical Physics Letters* **94**, 461–466 (1983).

9. G. B. Piland, J. J. Burdett, T.-Y. Hung, P.-H. Chen, C.-F. Lin, T.-L. Chiu, J.-H. Lee, C. J. Bardeen, Dynamics of molecular excitons near a semiconductor surface studied by fluorescence quenching of polycrystalline tetracene on silicon. *Chemical Physics Letters* **601**, 33–38 (2014).

10. B. Daiber, S. P. Pujari, S. Verboom, S. L. Luxembourg, S. W. Tabernig, M. H. Futscher, J. Lee, H. Zuilhof, B. Ehrler, A method to detect triplet exciton transfer from singlet fission materials into silicon solar cells: Comparing different surface treatments. *J. Chem. Phys.* **152**, 114201 (2020).

11. M. Einzinger, T. Wu, J. F. Kompalla, H. L. Smith, C. F. Perkinson, L. Nienhaus, S. Wieghold, D. N. Congreve, A. Kahn, M. G. Bawendi, M. A. Baldo, Sensitization of silicon by singlet exciton fission in tetracene. *Nature* **571**, 90–94 (2019).

12. Paper in preparation with authors N.N., Alexandra Alexiu, C.F.P., W.A.T., T.V.V., M.A.B.

13. J. Niederhausen, R. W. MacQueen, E. Özkol, C. Gersmann, M. H. Futscher, M. Liebhaber, D. Friedrich, M. Borgwardt, K. A. Mazzio, P. Amsalem, M. H. Nguyen, B. Daiber, M.

Mews, J. Rappich, F. Ruske, R. Eichberger, B. Ehrler, K. Lips, Energy-Level Alignment Tuning at Tetracene/c-Si Interfaces. *J. Phys. Chem. C* **124**, 27867–27881 (2020).

14. T. C. Wu, N. J. Thompson, D. N. Congreve, E. Hontz, S. R. Yost, T. Van Voorhis, M. A. Baldo, Singlet fission efficiency in tetracene-based organic solar cells. *Appl. Phys. Lett.* **104**, 193901 (2014).

15. S. Banerjee, M. K. Das, A review of Al2O3 as surface passivation material with relevant process technologies on c-Si solar cell. *Opt Quant Electron* **53**, 60 (2021).

16. W. Gao, A. Kahn, Electronic structure and current injection in zinc phthalocyanine doped with tetrafluorotetracyanoquinodimethane: Interface versus bulk effects. *Organic Electronics* **3**, 53–63 (2002).

17. P. S. Vincett, E. M. Voigt, K. E. Rieckhoff, Phosphorescence and Fluorescence of Phthalocyanines. *The Journal of Chemical Physics* **55**, 4131–4140 (1971).

18. P. J. Jadhav, A. Mohanty, J. Sussman, J. Lee, M. A. Baldo, Singlet Exciton Fission in Nanostructured Organic Solar Cells. *Nano Lett.* **11**, 1495–1498 (2011).

19. N. C. Giebink, G. P. Wiederrecht, M. R. Wasielewski, S. R. Forrest, Ideal diode equation for organic heterojunctions. I. Derivation and application. *Phys. Rev. B* **82**, 155305 (2010).

20. C. K. Renshaw, S. R. Forrest, Excited state and charge dynamics of hybrid organic/inorganic heterojunctions. I. Theory. *Phys. Rev. B* **90**, 045302 (2014).

21. H.-D. Um, K. Lee, I. Hwang, J. Park, D. Choi, N. Kim, H. Kim, K. Seo, Progress in silicon microwire solar cells. *J. Mater. Chem. A* **8**, 5395–5420 (2020).

22. K. Lee, I. Hwang, N. Kim, D. Choi, H.-D. Um, S. Kim, K. Seo, 17.6%-Efficient radial junction solar cells using silicon nano/micro hybrid structures. *Nanoscale* **8**, 14473–14479 (2016).

23. I. Hwang, H.-D. Um, B.-S. Kim, M. Wober, K. Seo, Flexible crystalline silicon radial junction photovoltaics with vertically aligned tapered microwires. *Energy Environ. Sci.* **11**, 641–647 (2018).

24. K. Lee, S. Shin, W. J. Lee, D. Choi, Y. Ahn, M. Park, D. Seo, K. Seo, Sunlight-Activatable ROS Generator for Cell Death Using TiO2/c-Si Microwires. *Nano Lett.* **21**, 6998–7004 (2021).

25. H.-D. Um, I. Hwang, N. Kim, Y. J. Yu, M. Wober, K.-H. Kim, K. Seo, Microgrid Electrode for Si Microwire Solar Cells with a Fill Factor of Over 80%. *Advanced Materials Interfaces* **2**, 1500347 (2015).

26. A. O. Adeyeye, G. Shimon, "Chapter 1 - Growth and Characterization of Magnetic Thin Film and Nanostructures" in *Handbook of Surface Science*, R. E. Camley, Z. Celinski, R. L. Stamps, Eds. (North-Holland, 2015) vol. 5 of *Magnetism of Surfaces, Interfaces, and Nanoscale Materials*, pp. 1–41.

27. G. F. Burkhard, E. T. Hoke, M. D. McGehee, Accounting for Interference, Scattering, and Electrode Absorption to Make Accurate Internal Quantum Efficiency Measurements in Organic and Other Thin Solar Cells. *Advanced Materials* **22**, 3293–3297 (2010).

28. S. Dong, A. Ong, C. Chi, Photochemistry of various acene based molecules. *Journal of Photochemistry and Photobiology C: Photochemistry Reviews* **38**, 27–46 (2019).

29. E. A. Kraut, R. W. Grant, J. R. Waldrop, S. P. Kowalczyk, Precise Determination of the Valence-Band Edge in X-Ray Photoemission Spectra: Application to Measurement of Semiconductor Interface Potentials. *Phys. Rev. Lett.* **44**, 1620–1623 (1980).

30. E. Epifanovsky, A. T. B. Gilbert, X. Feng, J. Lee, Y. Mao, N. Mardirossian, P. Pokhilko, A. F. White, M. P. Coons, A. L. Dempwolff, Z. Gan, D. Hait, P. R. Horn, L. D. Jacobson, I. Kaliman, J. Kussmann, A. W. Lange, K. U. Lao, D. S. Levine, J. Liu, S. C. McKenzie, A. F. Morrison, K. D. Nanda, F. Plasser, D. R. Rehn, M. L. Vidal, Z.-Q. You, Y. Zhu, B. Alam, B. J. Albrecht, A. Aldossary, E. Alguire, J. H. Andersen, V. Athavale, D. Barton, K. Begam, A. Behn, N. Bellonzi, Y. A. Bernard, E. J. Berquist, H. G. A. Burton, A. Carreras, K. Carter-Fenk, R. Chakraborty, A. D. Chien, K. D. Closser, V. Cofer-Shabica, S. Dasgupta, M. de Wergifosse, J. Deng, M. Diedenhofen, H. Do, S. Ehlert, P.-T. Fang, S. Fatehi, Q. Feng, T. Friedhoff, J. Gayvert, Q. Ge, G. Gidofalvi, M. Goldey, J. Gomes, C. E. González-Espinoza, S. Gulania, A. O. Gunina, M. W. D. Hanson-Heine, P. H. P. Harbach, A. Hauser, M. F. Herbst, M. Hernández Vera, M. Hodecker, Z. C. Holden, S. Houck, X. Huang, K. Hui, B. C. Huynh, M. Ivanov, Á. Jász, H. Ji, H. Jiang, B. Kaduk, S. Kähler, K. Khistyaev, J. Kim, G. Kis, P. Klunzinger, Z. Koczor-Benda, J. H. Koh, D. Kosenkov, L. Koulias, T. Kowalczyk, C. M. Krauter, K. Kue, A. Kunitsa, T. Kus, I. Ladjánszki, A. Landau, K. V. Lawler, D. Lefrancois, S. Lehtola, R. R. Li, Y.-P. Li, J. Liang, M. Liebenthal, H.-H. Lin, Y.-S. Lin, F. Liu, K.-Y. Liu, M. Loipersberger, A. Luenser, A. Manjanath, P. Manohar, E. Mansoor, S. F. Manzer, S.-P. Mao, A. V. Marenich, T. Markovich, S. Mason, S. A. Maurer, P. F. McLaughlin, M. F. S. J. Menger, J.-M. Mewes, S. A. Mewes, P. Morgante, J. W. Mullinax, K. J. Oosterbaan, G. Paran, A. C. Paul, S. K. Paul, F. Pavošević, Z. Pei, S. Prager, E. I. Proynov, Á. Rák, E. Ramos-Cordoba, B. Rana, A. E. Rask, A. Rettig, R. M. Richard, F. Rob, E. Rossomme, T. Scheele, M. Scheurer, M. Schneider, N. Sergueev, S. M. Sharada, W. Skomorowski, D. W. Small, C. J. Stein, Y.-C. Su, E. J. Sundstrom, Z. Tao, J. Thirman, G. J. Tornai, T. Tsuchimochi, N. M. Tubman, S. P. Veccham, O. Vydrov, J. Wenzel, J. Witte, A. Yamada, K. Yao, S. Yeganeh, S. R. Yost, A. Zech, I. Y. Zhang, X. Zhang, Y. Zhang, D. Zuev, A. Aspuru-Guzik, A. T. Bell, N. A. Besley, K. B. Bravaya, B. R. Brooks, D. Casanova, J.-D. Chai, S. Coriani, C. J. Cramer, G. Cserey, A. E. DePrince III, R. A. DiStasio Jr., A. Dreuw, B. D. Dunietz, T. R. Furlani, W. A. Goddard III, S. Hammes-Schiffer, T. Head-Gordon, W. J. Hehre, C.-P. Hsu, T.-C. Jagau, Y. Jung, A. Klamt, J. Kong, D. S. Lambrecht, W. Liang, N. J. Mayhall, C. W. McCurdy, J. B. Neaton, C. Ochsenfeld, J. A. Parkhill, R. Peverati, V. A. Rassolov, Y. Shao, L. V. Slipchenko, T. Stauch, R. P. Steele, J. E. Subotnik, A. J. W. Thom, A. Tkatchenko, D. G. Truhlar, T. Van Voorhis, T. A. Wesolowski, K. B. Whaley, H. L. Woodcock III, P. M. Zimmerman, S. Faraji, P. M. W. Gill, M. Head-Gordon, J. M. Herbert, A. I. Krylov, Software for the frontiers of quantum chemistry: An overview of developments in the Q-Chem 5 package. *The Journal of Chemical Physics* **155**, 084801 (2021).

31. C. Adamo, V. Barone, Toward reliable density functional methods without adjustable parameters: The PBE0 model. *The Journal of Chemical Physics* **110**, 6158–6170 (1999).

32. A. Schäfer, C. Huber, R. Ahlrichs, Fully optimized contracted Gaussian basis sets of triple zeta valence quality for atoms Li to Kr. *The Journal of Chemical Physics* **100**, 5829–5835 (1994).

33. R. Baer, E. Livshits, U. Salzner, Tuned Range-Separated Hybrids in Density Functional Theory. *Annual Review of Physical Chemistry* **61**, 85–109 (2010).

34. S. Hirata, M. Head-Gordon, Time-dependent density functional theory within the Tamm–Dancoff approximation. *Chemical Physics Letters* **314**, 291–299 (1999).

35. M. J. G. Peach, M. J. Williamson, D. J. Tozer, Influence of Triplet Instabilities in TDDFT. *J. Chem. Theory Comput.* **7**, 3578–3585 (2011).

36. B. Mennucci, R. Cammi, J. Tomasi, Excited states and solvatochromic shifts within a nonequilibrium solvation approach: A new formulation of the integral equation formalism method at the self-consistent field, configuration interaction, and multiconfiguration self-consistent field level. *The Journal of Chemical Physics* **109**, 2798–2807 (1998).

37. H. Sun, S. Ryno, C. Zhong, M. K. Ravva, Z. Sun, T. Körzdörfer, J.-L. Brédas, Ionization Energies, Electron Affinities, and Polarization Energies of Organic Molecular Crystals: Quantitative Estimations from a Polarizable Continuum Model (PCM)-Tuned Range-Separated Density Functional Approach. *J. Chem. Theory Comput.* **12**, 2906–2916 (2016).

38. D. Stahl, F. Maquin, Charge-stripping mass spectrometry of molecular ions from polyacenes and molecular orbital theory. *Chemical Physics Letters* **108**, 613–617 (1984).

39. J. Berkowitz, Photoelectron spectroscopy of phthalocyanine vapors. *The Journal of Chemical Physics* **70**, 2819–2828 (1979).

40. G. Man, J. Schwartz, J. C. Sturm, A. Kahn, Electronically Passivated Hole-Blocking Titanium Dioxide/Silicon Heterojunction for Hybrid Silicon Photovoltaics. *Advanced Materials Interfaces* **3**, 1600026 (2016).

**Acknowledgments:**

We thank UNIST Central Research Facilities (UCRF) for the support of its facilities and equipment.

**Funding:**

US Department of Energy, Office of Science, Office of Basic Energy Sciences, Materials Chemistry Program through award number DE-FG02-07ER46454 (NN, CFP, AL, LPW, TVV, MAB)

National Research Foundation of Korea NRF-2022R1C1C2012714 (KL)

National Science Foundation Graduate Research Fellowship Grant No. 1122374 (CFP)

Global Research Outreach Program of Samsung Advanced Institute of Technology (KL, CFP)

New Renewable Energy Core Technology Development Project of the Korea Institute of Energy Technology Evaluation and Planning (KETEP) grant No. 20223030010240 from the Ministry of Trade, Industry & Energy, Republic of Korea (YL, KS)

US Department of Energy, Office of Science, Office of Basic Energy Sciences, Physical Behavior of Materials Program award number DE-SC0019345 (WAT, NN)

US Department of Energy, Office of Science, Office of Basic Energy Sciences, Division of Materials Sciences and Engineering through award number DE-SC0012458 (XZ, SL, AK)

Lindemann Trust Fellowship of the English Speaking Union (TKB)

**Author contributions:**

Conceptualization: MAB, NN, KL, CFP

Fabrication of singlet fission top half of solar cells and device EQE measurement: NN, KL, AL

Device J-V measurements and SEM characterization: KL

Fabrication of silicon microwire solar cells: KL, YL

Ultraviolet photoelectron spectroscopy measurements: XZ, SL

Density functional theory calculations: LPW

Optical modeling simulations: NN

Project administration: MAB

Supervision: MGB, TVV, AK, WAT, KS, MAB

Writing – original draft: NN, MAB, KL

Writing – review and editing: NN, KL, CFP, AL, YL, XZ, SL, LPW, TKB, MGB, TVV, WAT, AK, KS, MAB

**Competing interests:** MIT has filed a US patent application for the two-part interlayer that names NN, CFP and MAB as inventors (Application No.: 18/689,764) and a provisional application for the device architecture that names NN, KL, AL, CFP and MAB as inventors (Serial No.: 63/596585).

**Data and materials availability:** All data are available in the main text or the supplementary materials.

**Supplementary Materials**

Materials and Methods

Supplementary Text

Figs. S1 to S10

Tables S1 to S5

References (*29–40*)

# Supplementary Materials for

## Exciton Fission Enhanced Silicon Solar Cell

Narumi Nagaya, Kangmin Lee, Collin F. Perkinson, Aaron Li, Youri Lee, Xinjue Zhong, Sujin Lee, Leah P. Weisburn, Tomi K. Baikie, Moungi G. Bawendi, Troy Van Voorhis, William A. Tisdale, Antoine Kahn, Kwanyong Seo, Marc A. Baldo
Corresponding author: baldo@mit.edu

**The PDF file includes:**

    Materials and Methods
    Supplementary Text
    Figs. S1 to S10
    Tables S1 to S5
    References (*29–40*)

**Materials and Methods**

Fabrication of tapered *c*-Si microwire arrays

The photoresist dot arrays (2 μm diameter, 1 μm spacing) were periodically patterned onto the *c*-Si wafer using AZ-nLOF-2035 photoresist (AZ Electronic Materials) via a photolithography process. The patterned *c*-Si wafer was then etched by deep reactive ion etching (DRIE, Tegal 200) with 1500 W source power, 100 W stage power, and 45 mTorr gas pressure using $SF_6$ (250 sccm) and $C_4F_8$ (150 sccm) as source gases. Following the DRIE process, cleaning was conducted using a piranha solution (3:1 mixture of sulfuric acid and hydrogen peroxide (30% w/w in $H_2O$)). To fabricate tapered *c*-Si microwires, the *c*-Si microwires were immersed in an $HF/HNO_3/CH_3COOH$ solution (RSE-100, Transene) for ~10 seconds.

Fabrication of tapered *c*-Si microwire solar cells

To fabricate $n^+$-p-type MWs, after the fabrication of the tapered p-type *c*-Si microwire array on the *c*-Si wafers, a back surface field (BSF) layer was formed in a tube furnace under a mixed atmosphere of $O_2$ (250 sccm) and $N_2$ (1000 sccm) at 980 °C using a boron spin-on-dopant source (B155, Filmtronics). Following the BSF layer formation, a $SiO_2$ layer was thermally grown on the wafer in a furnace. This $SiO_2$ layer was removed from the front of the substrate with a buffered oxide etchant (BOE) after the backside was protected with photoresist (AZ4330, AZ Electronic Materials). Subsequently, an additional 200-nm-thick $SiO_2$ diffusion barrier was deposited on the rear side of the wafer via plasma-enhanced chemical vapor deposition (PEH-600). Next, an emitter layer was formed in a tube furnace under a mixed atmosphere of $O_2$ (125 sccm) and $N_2$ (500 sccm) at 800 °C using a phosphorus spin-on-dopant source (P509, Filmtronics). Then, after removing the oxide layer using BOE, SC1 and SC2 cleaning processes were conducted. An $AlO_x$ passivation layer with a thickness of 1 nm was deposited on the front side of the wafer, while a 10-nm-thick $Al_2O_3$ layer was deposited on the rear side of the wafer via ALD (Lucida D100, NCD) and annealed under 500 sccm of a mixed gas of $Ar/H_2$ (v/v = 96:4) at 500 °C in a tube furnace. Patterns for the microgrid electrodes and localized back contact surface were formed on the front and rear sides of the wafer, respectively, via photolithography, using AZ4330 photoresist. For electrode fabrication, the wafer was immersed in a BOE solution, and 600 nm thick Al films were deposited on both sides of the wafer via thermal evaporation.

To fabricate $p^+$-n-type MW, after fabricating the tapered n-type *c*-Si microwire array on the *c*-Si wafers, the process was the same as for $n^+$-p-type MW, except for the BSF and emitter formation conditions. A BSF was formed in a tube furnace under a mixed atmosphere of $O_2$ (125 sccm) and $N_2$ (500 sccm) at 860 °C using a phosphorous spin-on-dopant source (P509, Filmtronics). Then, an emitter layer was formed on the front surface of the wafers under a mixed atmosphere of $O_2$ (250 sccm) and $N_2$ (1000 sccm) at 880 °C using a boron spin-on-dopant source (B155, Filmtronics).

Singlet fission top-side fabrication

Before deposition, the samples were dipped in acetone for 5 minutes, and then in IPA for 5 minutes at a low sonication power. Afterward, the samples were dried with $N_2$ gas.

After cleaning, the samples were transferred into a nitrogen glovebox for deposition of the organic layers. ZnPc was purchased from Luminescence Technology Corp (sublimed grade,

>99% purity) and used as received. Tc was purchased from Sigma Aldrich (sublimed grade, 99.99% trace metal basis) and purified once through a sublimation and recondensation process in a three-zone tube furnace before deposition. The organic layers were deposited by thermal evaporation in a vacuum chamber at a pressure of $< 1 \times 10^{-6}$ torr. 1.5 nm of ZnPc was deposited at a 0.2 Å/s rate, followed by 30 nm of Tc deposited at a 1 Å/s rate. Thickness calibrations were obtained from ellipsometry measurements performed on planar substrates. The surface area of the microwire substrates is much larger. Thus, the true area density of ZnPc (1.5 nm nominal thickness) and Tc (30 nm nominal thickness) in the microwire devices is expected to be lower.

The samples were then encapsulated in a dry nitrogen atmosphere (<1 ppm $O_2$) with a glass slide and ultraviolet curable epoxy. A small piece of foil was placed to protect the active area during the UV exposure for the curing step.

Photoelectron spectroscopy
Tc and ZnPc layers were deposited on $AlO_x$-silicon surfaces by thermal evaporation in an ultrahigh-vacuum chamber ($10^{-9}$ Torr) connected to the analysis chamber. 1.5 nm of ZnPc was deposited at a rate of about 0.1 Å/s, followed by 20 nm of Tc at a deposition rate of about 1 Å/s. Then, ultra-violet and X-ray photoemission spectroscopy (UPS, XPS) measurements were conducted under ultrahigh vacuum conditions ($10^{-10}$ Torr) at room temperature. UPS utilized He I photons (21.22 eV) emitted by a discharged lamp to determine the work function of the sample and the energy position of the ZnPc and Tc HOMO with respect to the Fermi level. The experimental energy resolution for UPS is typically 0.10 eV. The position of the Si valence band maximum ($E_V$), hidden by the organic and $AlO_x$ overlayers in UPS, was determined from the position of the Si 2p core level following the Kraut method (*29*). XPS was performed using non-monochromatized Al Kα X-rays (1486.70 eV). The Fermi level reference for all measurements was determined on a clean Au surface.

Device measurements
The current density–voltage curve of the fabricated solar cells was measured using a Newport Orel class A solar simulator (Model 91159) under AM 1.5G illumination. The incident flux was calibrated using a calibrated reference Si solar cell certified by Newport (Model 91150-KG5). The solar cells were measured from −1.0 to 1.0 V at a temperature of 25 °C in air. External quantum efficiency measurements were performed using a 150 W xenon lamp coupled to a Newport monochromator. The light output from the monochromator was mechanically chopped at a frequency of 330 Hz and the photocurrent from the device was measured through a lock-in amplifier. The light input intensity was measured using a Newport photodetector, responsivity calibrated by Newport. The surface morphology of the solar cells was characterized by SEM (Zeiss Merlin High-resolution SEM).

Modeling
Full details on the modeling performed to fit the differential EQE are presented in the Supplementary Text. A transfer-matrix method (*27*) was used to simulate the absorption of light through the device stack. The fits were performed using custom MATLAB code.

**Supplementary Text**

Tetracene and zinc phthalocyanine calculations

The triplet excited state energies and the highest occupied molecular orbital (HOMO) energies for tetracene (Tc) and zinc phthalocyanine (ZnPc) molecules are calculated using several methods with Q-Chem 5.4 (*30*). The structures of the molecules are optimized with the standard hybrid PBE0 functional and def2-TZVP basis set in gas phase (*31*, *32*).

The PBE0 functional is used and compared to tuned range-separated functionals, LRC- ωPBE, to better evaluate the molecular orbital energies. These functionals are tuned uniquely for each molecule per Koopman's theorem to ensure consistency between ionization and frontier orbital energies (*33*). This tuning matches the ionization energy for the ground and anionic states to each's negative HOMO energy (Eq. 1).

$$E(A^+) - E(A) = -E(HOMO_A) \qquad (1)$$

These functionals are tuned for gas phase molecules without solvent.

Time-dependent density functional theory (TDDFT) with the Tann-Dancoff Approximation (TDA) is used to examine the excited triplets for the molecules. This method is shown, and in this case proven, to be less susceptible to triplet instability issues than TDDFT (*34*, *35*). ΔSCF energies are also calculated for the triplet excitations on both molecules for comparison. The calculations are performed with and without implicit polarizable continuum model solvent in the integral equation formalism (IEF-PCM) (*36*). The IEF-PCM solvent is given a weak dielectric of 3.0 to roughly simulate molecules in a thin, organic film, as in the experimental solar cell. To the best of our knowledge, the dielectric constants for these films are not reported, so we use a conservative constant in the lower range for pentacene and other similar systems (*37*). Regardless, we find changing this constant has little influence on the results.

The TDA triplet, ΔSCF triplet, ground state HOMO, and ionization energies for each functional for both molecules without IER-PCM solvent are given in eV in Table S3. The same data, with IEF-PCM solvent correction, is given in Table S4.

Here, the tuned functionals for each molecule show agreement between the ionization energy and HOMO energy levels, so we use these functionals to approximate the HOMO energy delta between Tc and ZnPc. We find LRC- ωPBE$_{207}$ for Tc yielded an ionization energy of 6.7 eV, while LRC- ωPBE$_{156}$ gives an ionization energy of 6.3 eV for ZnPc. These are in good agreement with established experimental values for the gas phase molecules (*38*, *39*).

In applying IEF-PCM corrections, we aim to better simulate the molecules in a thin film rather than in the gas phase. We take the ionization energy to be the ground state HOMO for each of the molecules, which is a better proxy for the frontier orbital energies and ordering with solvent corrections. Using these ionization energies for tetracene with LRC- ωPBE$_{207}$ (5.7 eV) and ZnPc with LRC- ωPBE$_{156}$ (5.5 eV), we estimate the difference in the HOMO energies to be approximately 0.2 eV. This deeper Tc HOMO level agrees with experimental XPS data and confirms the molecular difference and ordering, even when the compounds are not deposited on the silicon cell.

We also find that the triplet energies are affected little by the IEF-PCM corrections from the TDA and ΔSCF calculations. We estimate that, through this survey of methods, the tetracene $T_1$ energy is about 1.4 eV, while the ZnPc $T_1$ energy is about 1.2 eV. Again, this aligns with experimental values and relative ordering for the two molecules within errors of density functional theory.

Ultraviolet and X-ray photoelectron spectroscopy (UPS & XPS) measurements

UPS and XPS are used to determine the energy alignments of Tc/ZnPc/AlO$_x$/Si. The energy difference between the Si2p core level ($E_{Si2p3/2}$) and the valence band edge ($E_v$) is calibrated using an argon-ion sputtered H-passivated Si surface. As shown in Figure S7, UPS measures $E_v$(Si) located at 0.4 eV below the Fermi level ($E_F$), and XPS measures $E_{Si2p3/2}$(Si) at 99.35 eV below $E_F$, which gives $E_{Si2p3/2}$(Si) - $E_v$(Si) = 99.35 – 0.4 = 98.95 eV. This energy difference is comparable to values in literature (*40*). Therefore, for n$^+$-doped *c*-Si, $E_v$ is at 99.91 eV – 98.95 eV = 0.96 eV below $E_F$, and for p$^+$-doped *c*-Si, at 98.95 eV – 98.95 eV ~ 0 eV, within experimental resolution at the Fermi level.

Optical modeling and tetracene sensitization efficiency calculations

The absorption of light by each of the layers in the complete device stack and the total reflection is simulated using the transfer matrix method[12]. The optical absorption of each layer in the Tc/ZnPc/n$^+$-p device is calculated for a stack with 30 nm of tetracene ($Abs_{Tc,Tc=30}$), 1.5 nm of zinc phthalocyanine ($Abs_{ZnPc,Tc=30}$), 1 nm of Al$_2$O$_3$ ($Abs_{Al_2O_3,Tc=30}$), and 180 µm of silicon ($Abs_{Si,Tc=30}$). As the control, we also calculate the optical absorption in each layer of an unsensitized n$^+$-p device consisting of 1 nm of Al$_2$O$_3$ ($Abs_{Al_2O_3,Tc=0}$), and 180 µm of silicon ($Abs_{Si,Tc=0}$).

To obtain the efficiency of tetracene sensitization, we analyze the percentage differential external quantum efficiency (EQE) ($\Delta EQE/EQE_0$) that results from coupling to tetracene. We define $\frac{\Delta EQE}{EQE_0} = \frac{EQE_{Tc}-EQE_0}{EQE_0}$, where $EQE_0$ is the silicon cell EQE before Tc and ZnPc deposition and $EQE_{Tc}$ is the silicon cell EQE after Tc and ZnPc deposition.

Microwire cells are especially effective at absorbing incident light[13]. Consequently, we assume that total device reflection losses are negligible in the microwire cells such that device EQEs are approximately equivalent to the device internal quantum efficiencies (IQEs): $\frac{\Delta EQE}{EQE_0} \approx \frac{\Delta IQE}{IQE_0}$ with the subscripts as defined previously.

We assume $IQE_0$ can be described by:
$$IQE_0(\lambda) = \eta_{Si,Tc=0}(\lambda)\phi_{Si,Tc=0}(\lambda)$$
where $\eta_{Si,Tc=0}$ is the internal quantum efficiency of the silicon without the deposited organic layers and $\phi_{Si,Tc=0}$ is the simulated fractional absorption of the silicon layer relative to the total device stack absorption without the deposited organic layers, $\phi_{Si,Tc=0} = \frac{Abs_{Si,Tc=0}}{\sum Abs_{l,Tc=0}}$. All optical absorption simulations are performed in a planar geometry using optical constants experimentally determined in the planar geometry.

Next, we assume that $IQE_{Tc}$ can be described by:

$$IQE_{Tc}(\lambda) = \eta_{Si,Tc=30}(\lambda)\phi_{Si,Tc=30}(\lambda) + \eta_{Tc}\phi_{Tc,Tc=30}(\lambda),$$

where $\eta_{Si,Tc=30}(\lambda)$ is the internal quantum efficiency of the silicon with the deposited organic layers, $\eta_{Tc}$ is the tetracene sensitization efficiency (ranging from 0 to 200%, encompassing efficiency of singlet fission, triplet exciton transport, triplet exciton transfer, and extraction of transferred carriers to silicon), $\phi_{Si,Tc=30}$ and $\phi_{Tc,Tc=30}$ are the simulated fractional absorption of the silicon and tetracene layers, respectively. We treat the tetracene contribution to the IQE as a separate additional term, because we expect the extraction of carriers absorbed by the silicon cell directly to differ from the extraction of transferred carriers from tetracene.

Then, we arrive at the expression:

$$\frac{\Delta EQE(\lambda)}{EQE_0(\lambda)} \approx \left(\frac{\eta_{Si,Tc=30}(\lambda)\phi_{Si,Tc=30}(\lambda)}{\eta_{Si,Tc=0}(\lambda)\phi_{Si,Tc=0}(\lambda)} + \frac{\eta_{Tc}\phi_{Tc,Tc=30}(\lambda)}{\eta_{Si,Tc=0}(\lambda)\phi_{Si,Tc=0}(\lambda)}\right) - 1.$$

Since we are interested in the relative effect of adding the organic layers to the device stack, we set $\eta_{Si,Tc=0} = 1$. Then, to obtain a lower bound for $\eta_{Tc}$, we also assume $\eta_{Si,Tc=30}(\lambda = 520\ nm) = 1$ (*i.e.* that the deposition of organic layers does not affect the passivation quality of the silicon surface). Finally, we fit for $\eta_{Tc}$ at λ = 520 nm. The simulated $\frac{\Delta EQE(\lambda)}{EQE_0(\lambda)}$ is presented for different $\eta_{Tc}$ values in the main text Fig. 4c, showing an optimal fit for $\eta_{Tc} = 138\%$. As noted in the Materials and Methods section, the material optical constants and experimental thickness calibrations are obtained from ellipsometry measurements performed on planar substrates. Consequently, the true area density of ZnPc (1.5 nm nominal thickness) and Tc (30 nm nominal thickness) in the microwire devices is substantially lower, leading to an overestimate of $\phi_{Tc,Tc=30}$ and an interpretation of fits for $\eta_{Tc}$ as a lower bound.

A similar analysis can be carried out for the other microwire device structures that are studied in this work. Table S5 summarizes the obtained fit parameters for $\eta_{Tc}$ in different device structures. Uncertainty ranges are calculated from the maximum and minimum fit values for $\eta_{Tc}$ across the tetracene absorption region (λ = 415 – 545 nm).

We observe that the assumption of a wavelength-independent internal quantum efficiency for silicon ($\frac{\eta_{Si,Tc=30}(\lambda)}{\eta_{Si,Tc=0}(\lambda)} = 1$) only yields good fits for the n$^+$-p Si MW (1 nm AlO$_x$) devices. Allowing $\frac{\eta_{Si,Tc=30}(\lambda)}{\eta_{Si,Tc=0}(\lambda)}$ to vary results in the following values for $\frac{\eta_{Si,Tc=30}(\lambda)}{\eta_{Si,Tc=0}(\lambda)}$, presented in Fig. S10. The apparent reduction in silicon quantum yield at shorter wavelengths is consistent with a deterioration in silicon passivation. This suggests that adding the organic layers may degrade the passivation quality for the devices that do not show transfer.

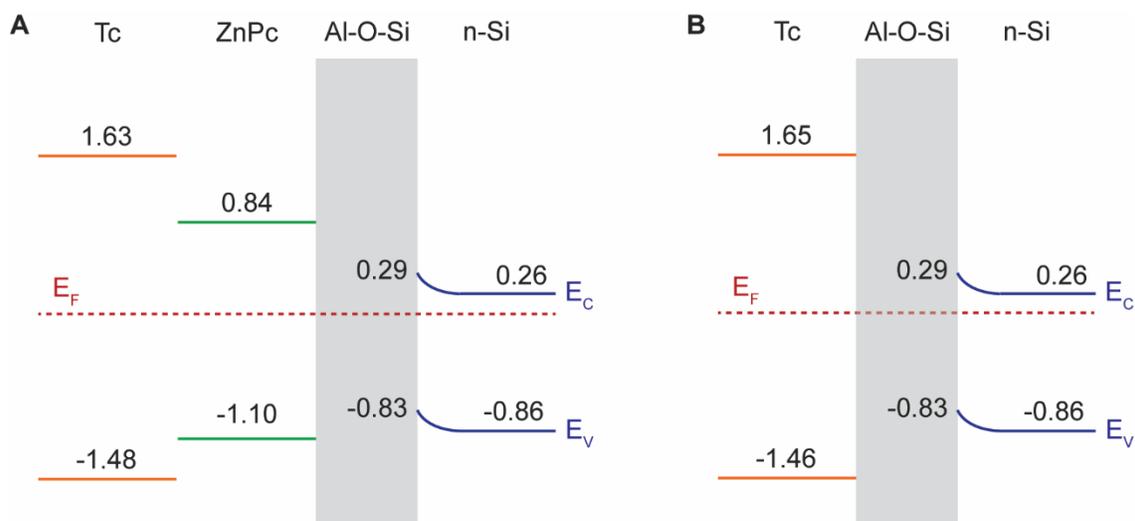

**Fig. S1.**
**Energy alignment summary for tetracene and zinc phthalocyanine on n-doped silicon.**
Summary of band alignments for (**A**) tetracene (Tc), zinc phthalocyanine (ZnPc) and $AlO_x$ deposited on an n-doped silicon surface (n-Si), and (**B**) tetracene (Tc) and $AlO_x$ deposited on the same n-doped silicon surface (n-Si). The red-dashed line indicates the position of the Fermi level of the system. The valence band maximum and highest occupied molecular orbital (HOMO) positions were measured using ultraviolet photoelectron spectroscopy. The conduction band minimum and lowest unoccupied molecular orbital (LUMO) positions were calculated using the electronic band gaps from previous UPS and inverse photoemission spectroscopy measurements (*11*, *16*). The bulk values for the conduction band ($E_C$) and valence band ($E_V$) of the doped silicon were calculated from the doping concentration. Energy values are in eV.

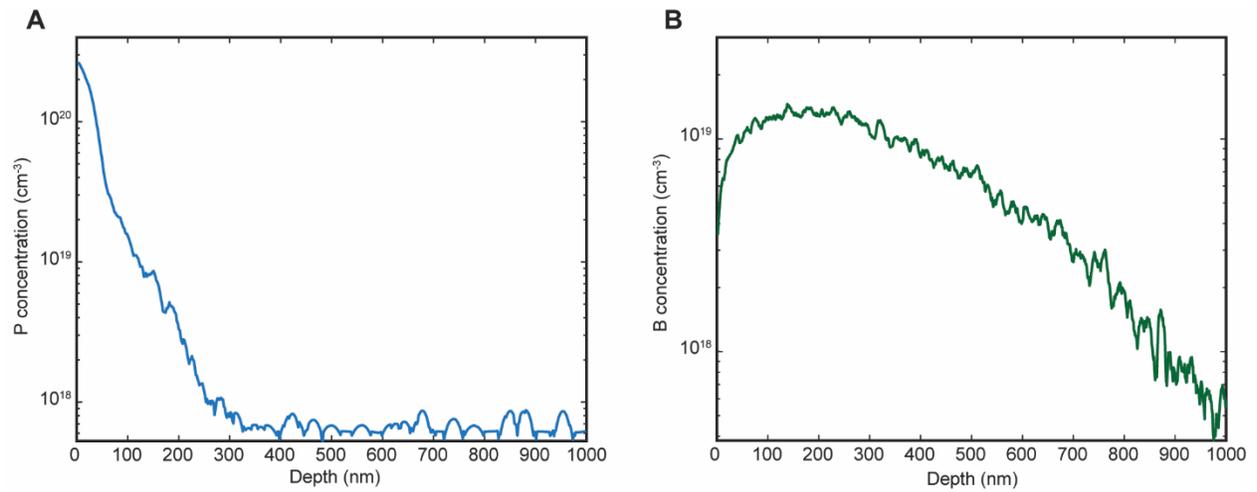

**Fig. S2.**
**SIMS depth profiles of $n^+$-p Si.** Time-of-flight secondary ion mass spectrometry (ToF-SIMS) depth profiles of $n^+$-p Si. (**A**) The junction depth of the emitter region ($n^+$) is approximately 300 nm. (**B**) The junction depth of the back surface field region ($p^+$) is approximately 1000 nm.

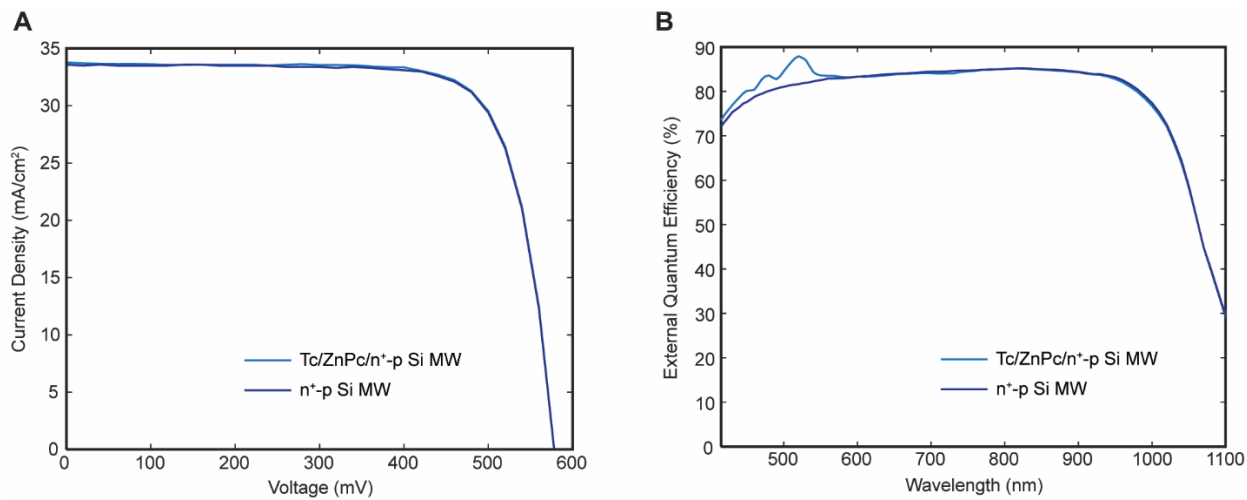

**Fig. S3.**
**Photovoltaic Performance of n$^+$-p Si MW devices.** (**A**) *J-V* characteristics of the n$^+$-p Si MW with and without Tc/ZnPc when illuminated under AM 1.5 G spectral conditions at 25 °C. (**B**) Full-scale external quantum efficiency spectra (EQE) of the n$^+$-p Si MW with and without Tc/ZnPc.

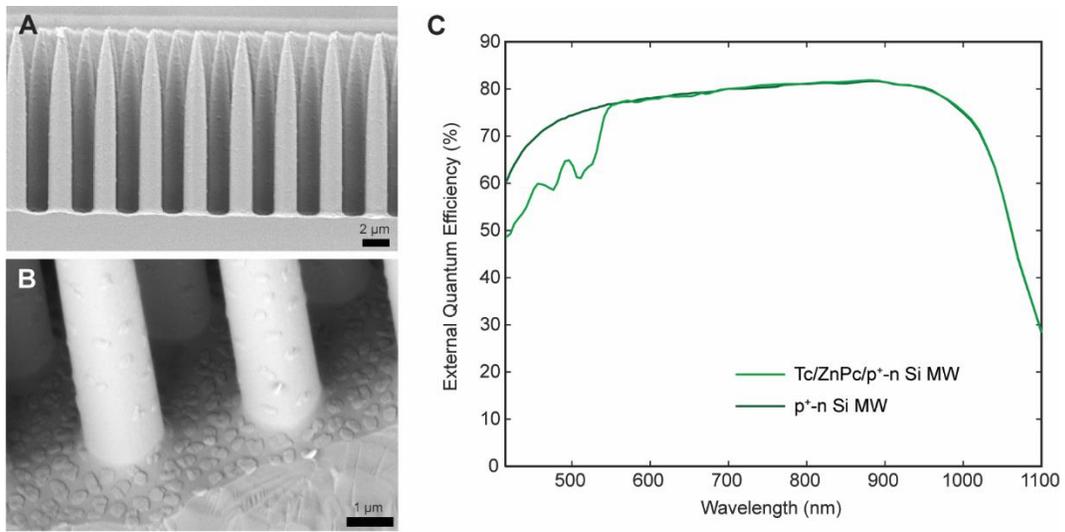

**Fig. S4.**
**Characterization of p$^+$-n Si MW devices.** (**A-B**) SEM images of the Tc/ZnPc/p$^+$-n Si MW, and (**C**) full scale EQE of the p$^+$-n Si MW with and without Tc/ZnPc.

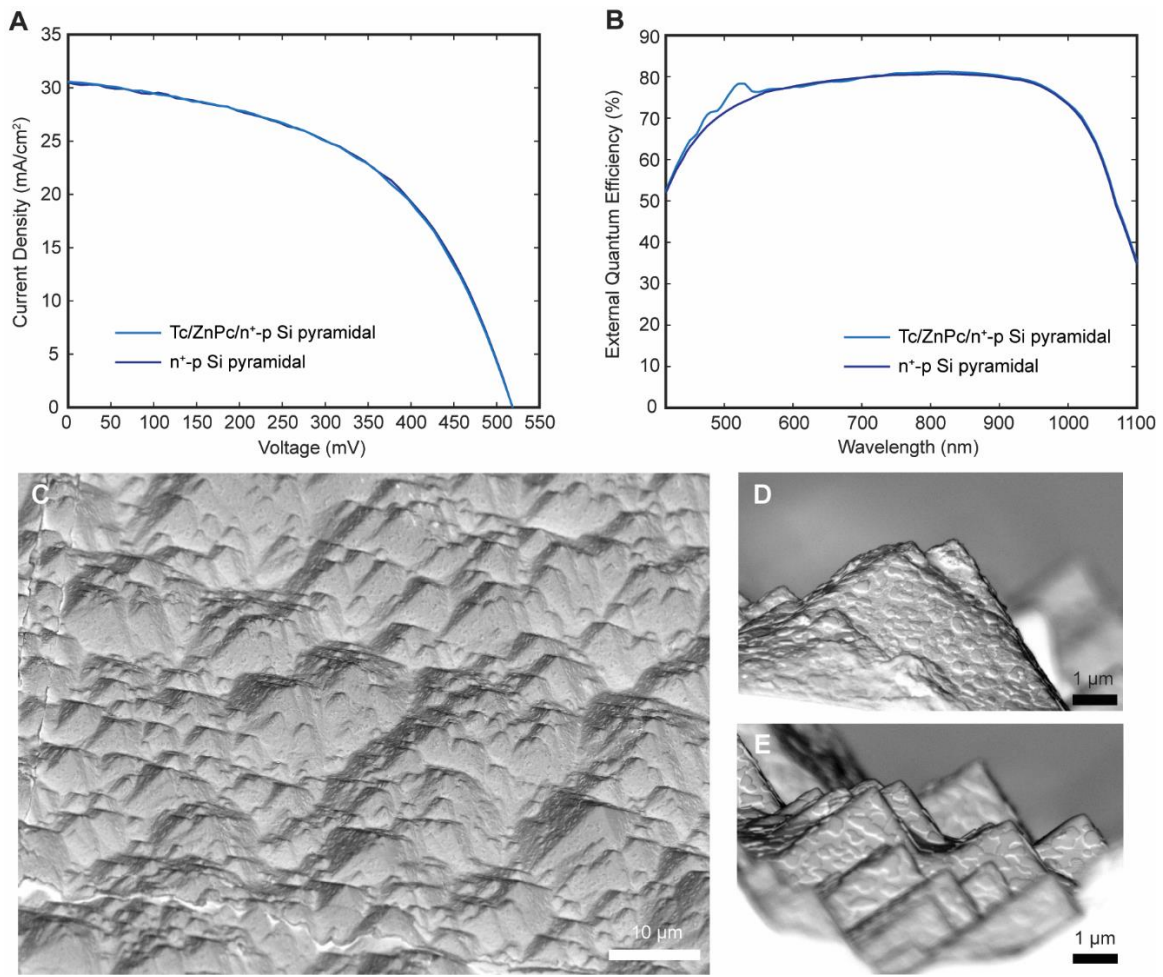

**Fig. S5.**

**Pyramidal-textured planar n$^+$-p Si device characterization.** (**A**) *J-V* characteristics of the n$^+$-p Si pyramidal cells with and without Tc/ZnPc when illuminated under AM 1.5 G spectral conditions at 25 °C. (**B**) External quantum efficiency spectra (EQE) of the n$^+$-p Si MW with and without Tc/ZnPc. (**C-E**) SEM images of the Tc/ZnPc/n$^+$-p Si pyramidal cells.

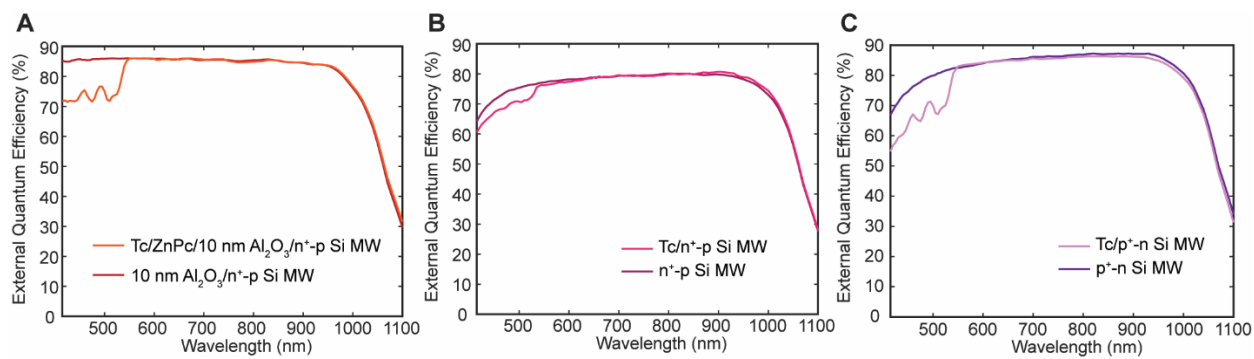

**Fig. S6.**

**Control silicon microwire device characterization.** (**A**) EQE spectra of $n^+$-p Si MW with 10 nm $Al_2O_3$ with and without Tc/ZnPc. (**B**) EQE spectra of $n^+$-p Si MW with and without Tc. (**C**) EQE spectra of $p^+$-n Si MW with and without Tc.

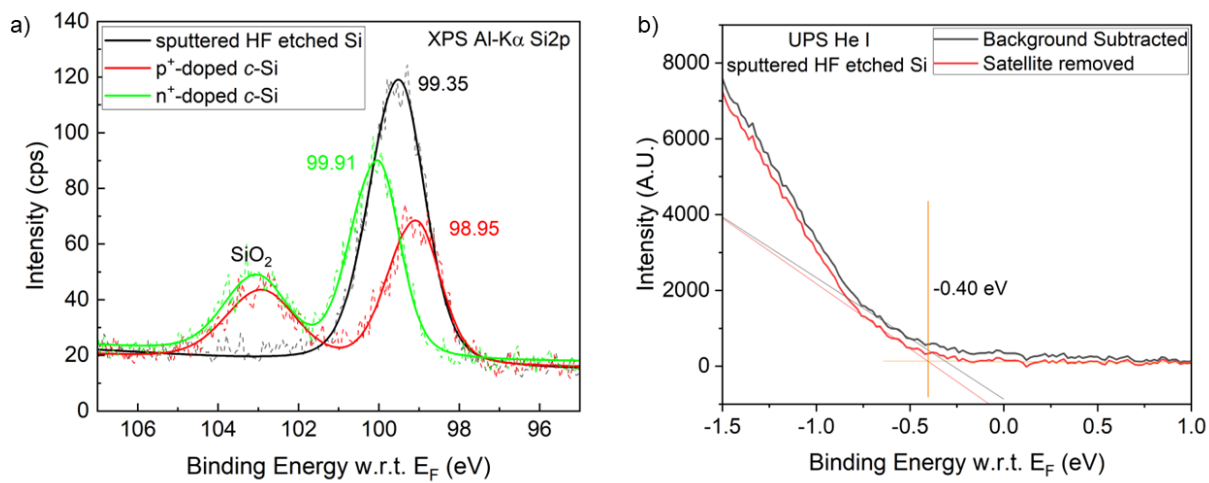

**Fig. S7.**
(**A**) High energy resolution XPS spectra of Si2p. (**B**) UPS valence band spectrum on argon-ion sputtered H-passivated Si surface.

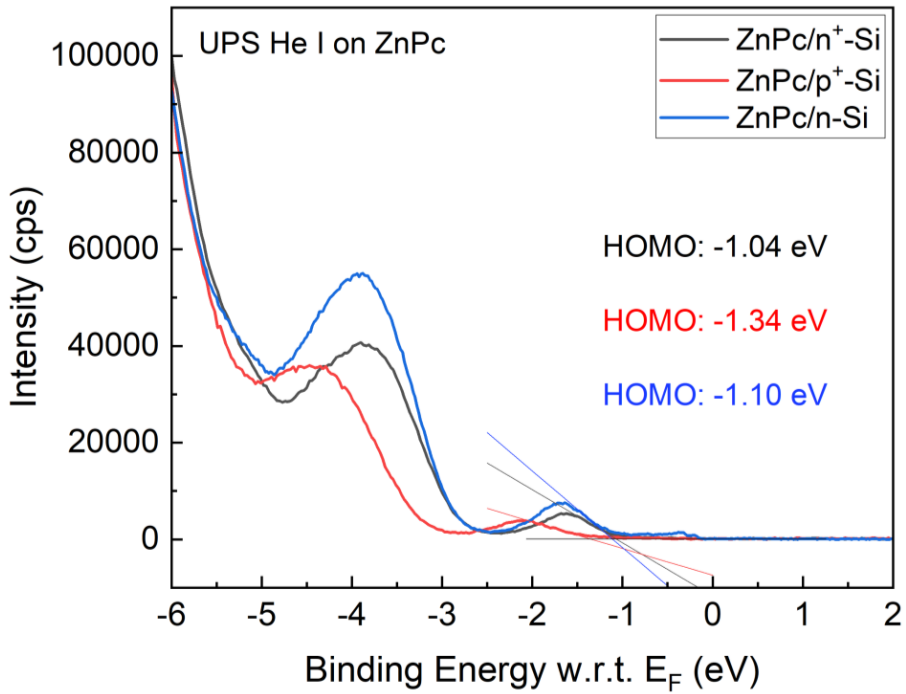

**Fig. S8.**

ZnPc HOMO edge position: UPS He I measurements on ZnPc on a highly n-doped silicon surface ($n^+$-Si), a highly p-doped silicon surface ($p^+$-Si), and an n-doped silicon surface (n-Si), respectively.

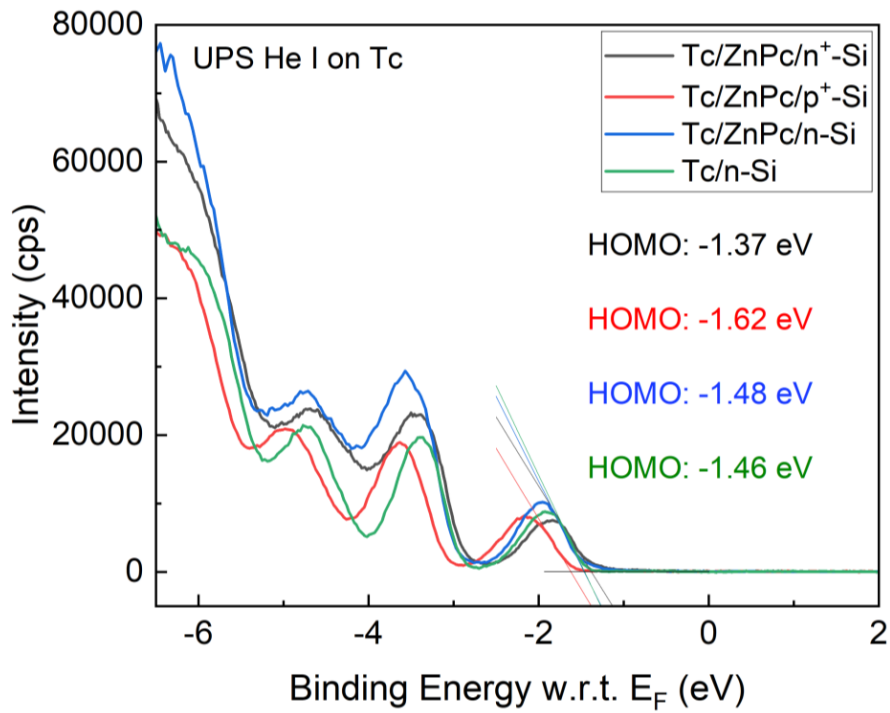

**Fig. S9.**

Tc HOMO edge position: UPS He I measurements on Tc/ZnPc/n$^+$-Si, Tc/ZnPc/p$^+$-Si, Tc/ZnPc/n-Si, and Tc/n-Si without ZnPc.

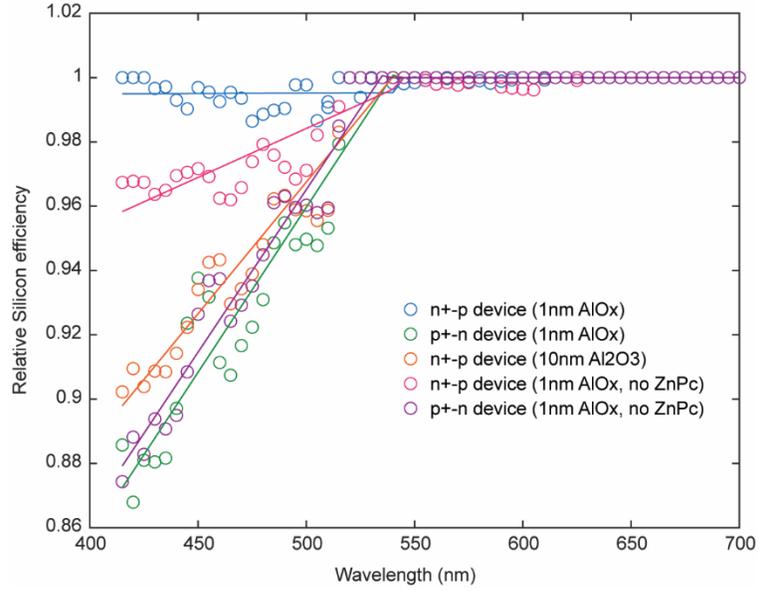

**Fig. S10.**

The circles show the relative silicon efficiencies $\frac{\eta_{Si,Tc=30}(\lambda)}{\eta_{Si,Tc=0}(\lambda)}$ obtained from fitting the model to the data and setting $\eta_{Tc}$ to a fixed value presented in Table S5. A piecewise linear fit from $\lambda = 415 - 545$ nm, and from $\lambda = 550 - 700$ nm is used to obtain the lines.

**Table S1.**

Photovoltaic properties of n$^+$-p Si MW solar cells with and without Tc/ZnPc

| Microwire | $J_{sc}$ (mA/cm$^2$) | $V_{oc}$ (mV) | FF (%) | PCE (%) |
|---|---|---|---|---|
| **n$^+$-p Si** | 33.57 | 578.1 | 77.05 | 14.95 |
| **Tc/ZnPc/n$^+$-p Si** | 33.79 | 578.0 | 76.85 | 15.01 |

**Table S2.**

Photovoltaic properties of pyramidal-textured planar $n^+$-p Si device with and without Tc/ZnPc

| Pyramidal-textured planar | $J_{sc}$ (mA/cm$^2$) | $V_{oc}$ (mV) | FF (%) | PCE (%) |
|---|---|---|---|---|
| $n^+$-p Si | 30.50 | 518.6 | 50.81 | 8.0 |
| Tc/ZnPc/$n^+$-p Si | 30.61 | 518.7 | 50.52 | 8.0 |

**Table S3.**

TDA triplet, ΔSCF triplet, ground state HOMO, and ionization energies for each functional for both molecules without IER-PCM solvent, given in eV

| (eV) | Tetracene | | | | Zinc Phthalocyanine | | | |
|---|---|---|---|---|---|---|---|---|
| Functional | TDA T1 | ΔSCF T1 | Ground State HOMO | Ionization Energy | TDA T1 | ΔSCF T1 | Ground State HOMO | Ionization Energy |
| PBE0 | 1.44 | 1.45 | -5.34 | 6.76 | 1.24 | 1.18 | -5.33 | 6.26 |
| LRC-$\omega$PBE$_{156}$: ZnPc Optimized | 1.51 | 1.55 | -6.45 | 6.70 | 1.25 | 1.18 | -6.29 | 6.29 |
| LRC-$\omega$PBE$_{207}$: Tc Optimized | 1.55 | 1.58 | -6.76 | 6.77 | 1.22 | 1.05 | -6.43 | 6.24 |

**Table S4.**

TDA triplet, ΔSCF triplet, ground state HOMO, and ionization energies for each functional for both molecules with IER-PCM solvent correction, given in eV.

| (eV) | Tetracene | | | | Zinc Phthalocyanine | | | |
|---|---|---|---|---|---|---|---|---|
| Functional | TDA T1 | ΔSCF T1 | Ground State HOMO | Ionization Energy | TDA T1 | ΔSCF T1 | Ground State HOMO | Ionization Energy |
| PBE0 | 1.44 | 1.45 | -5.33 | 5.54 | 1.23 | 1.17 | -5.33 | 5.44 |
| LRC-$\omega$PBE$_{156}$: ZnPc Optimized | 1.51 | 1.54 | -6.44 | 5.60 | 1.22 | 1.16 | -6.28 | 5.45 |
| LRC-$\omega$PBE$_{207}$: Tc Optimized | 1.55 | 1.58 | -6.75 | 5.67 | 1.15 | 1.04 | -6.42 | 5.40 |

**Table S5.**

Fitted tetracene sensitization efficiency values for each microwire device structure investigated in this work.

| Device structure | $\eta_{Tc}$ (%) |
|---|---|
| $n^+$-p Si MW (1 nm AlO$_x$) | 138 ± 6 |
| $p^+$-n Si MW (1 nm AlO$_x$) | 25 ± 42 |
| $n^+$-p Si MW (10 nm Al$_2$O$_3$) | 47 ± 45 |
| $n^+$-p Si MW (1 nm AlO$_x$, no ZnPc) | 74 ± 28 |
| $p^+$-n Si MW (1 nm AlO$_x$, no ZnPc) | 28 ± 43 |